\documentclass[epjCONF,columns]{svjour} 
\usepackage{graphics}
\usepackage[varg]{txfonts} 
\usepackage[latin1]{inputenc}
\session-title{Hadron Collider Physics symposium (HCP-2011)}
\begin{document}
\title{Searches for diboson production at the Tevatron\\ in final states containing heavy-flavor jets}
\author{Jean-Fran\c{c}ois Grivaz\thanks{\email{grivaz@lal.in2p3.fr}}}
\institute{Laboratoire de l'Accélérateur Linéaire (Orsay, France)\\
\\
On behalf of the CDF and D0 Collaborations.}
\abstract{
Recent searches performed by the CDF and D0 collaborations at the Tevatron for diboson production 
in final states containing heavy-flavor jets are reported. The searches for $WZ$ and $ZZ$ can be
regarded as the ultimate benchmark for the corresponding searches for a low-mass Higgs boson in the
$WH$ and $ZH$ final states. Using the exact same techniques as for those Higgs boson searches, the
D0 collaboration measured a cross section for $WZ/ZZ$ production of $1.13 \pm 0.36$ times its 
expectation in the standard model, with a diboson signal significance of 3.3 standard deviations
(2.9 expected).
} 
\maketitle
\section{Introduction}
\label{intro}
Diboson production in $p\bar{p}$ (or $pp$) collisions at high energy is a topic of interest 
in its own right. Since new physics could manifest itself differently in different final 
states, it is important that corresponding analyses be performed. At the Tevatron, diboson
production has been observed first in fully leptonic and later in semileptonic final 
states~\cite{Robson}.
With the large integrated luminosity now available, together with improved analysis techniques, it
has become possible to study diboson production in final states containing heavy-flavor (HF) jets.
An additional interest for this topic is that such final states are encountered almost identically
for a low-mass Higgs boson produced in association with a vector boson. It is this aspect of 
diboson production at the Tevatron that is the focus of this presentation.

The searches for a low-mass Higgs boson in $p\bar{p}$ collisions have already been reviewed at 
this conference~\cite{Sforza}. Therefore, only a brief summary of the methodology is given here. 
The relevant production modes are $W(\to\ell\nu)H$, $Z(\to\ell\ell)H$, and $Z(\to\nu\nu)H$,
with $\ell$ an electron or a muon and $H\to bb$. 
After a trigger selection based on isolated leptons and/or missing transverse energy (as expected
from neutrinos in the final state), simple kinematic and lepton-identification criteria are 
applied to reduce the initial sample to a manageable level without significant signal loss; 
the bulk of the remaining background originating from 
multijet (MJ) production is removed either by further kinematic criteria, or, more frequently in 
recent searches, by making use of some multivariate analysis technique (MVA); the sample is then 
enriched in HF jets through the application of a ``$b$-tagging'' algorithm, and often
divided into sub-samples according to the number of $b$-tagged jets and to the level of $b$ purity 
of those tagged jets; final discriminants are constructed for each of these sub-samples, in which
advantage is taken
of the kinematic differences between the signal and the remaining backgrounds, most importantly
$(W/Z)bb$, which is irreducible. A variety of MVAs are used, chosen based on some
optimization of the search sensitivity but also on the analyzers' expertise: Artificial Neural 
Networks (NN), Support Vector Machines (SVM), Boosted Decision Trees (BDT), Random Forests (RF).
The final discriminants are subjected to a statistical analysis based on the Log-Likelihood Ratio
(LLR) between the back\-ground-only and the signal+back\-ground hypotheses, with marginalization
over all nuisance parameter priors. 

This whole sequence of analysis procedures, designed to reach sensitivity to a tiny Higgs boson
signal in the presence of large backgrounds, and therefore apparently quite involved, 
would largely benefit in terms of reliability from being validated through the observation of a 
similar, but known signal. The search for diboson production with $Z\to bb$ in the final state
fulfills these requirements. Consider a Higgs boson with a mass of 115~GeV. In the final states 
of interest, the production cross sections in $p\bar{p}$ collisions at 1.96~TeV are:
\begin{itemize}
\item 27 fb for $WH\to\ell\nu bb$ with $\ell = e$ or $\mu$,
\item 5 fb for $ZH\to\ell\ell bb$ with $\ell\ell = ee$ or $\mu\mu$,
\item 15 fb for $ZH\to\nu\nu bb$,
\end{itemize}
for a total of 46~fb. Replacing $H\to bb$ by $Z\to bb$, the corresponding cross sections are
\begin{itemize}
\item 105 fb for $WZ\to\ell\nu bb$ with $\ell = e$ or $\mu$,
\item 24 fb for $ZZ\to\ell\ell bb$ with $\ell\ell = ee$ or $\mu\mu$,
\item 73 fb for $ZZ\to\nu\nu bb$,
\end{itemize}
for a total of 202~fb. These cross sections for diboson production are seen to be about 4.5 times
larger than the corresponding ones for a 115~GeV mass Higgs boson. It should however be kept in 
mind that the dijet mass resolution of the CDF and D0 detectors is not sufficient to separate the
$W$ and $Z$ dijet mass peaks, so that $WW\to\ell\nu cs$ is a significant resonant background. 
Furthermore, the non-resonant $(W/Z)bb$ and $(W/Z)cc$ backgrounds and their related systematic 
uncertainties are substantially larger than for a Higgs boson with a mass 25~GeV above that of 
the $Z$ boson. 
On the other hand, there is relatively more signal contribution from $Z\to cc$ than from
$H\to cc$. Altogether, the observation of $(W/Z)(Z\to bb)$, using the same techniques as in the
searches for a low-mass Higgs boson, can be considered as the ultimate benchmark for those 
searches at the Tevatron.

A total of eight conference notes or publications relevant to diboson production at the Tevatron 
with HF jets in the final state were made available for this conference by the CDF 
and D0 collaborations. The analyses use integrated luminosities ranging from 4.3 to 8.4~fb$^{-1}$. 
For reference, the cross sections used by the Tevatron Higgs Combination Working Group for inclusive
diboson production are: 11.34~pb for $WW$, 3.22~pb for $WZ$, and 1.20~pb for $ZZ$; they were 
obtained with {\sc mcfm}~\cite{MCFM} at next-to-leading order.
 
\section{Towards the benchmark}
\label{sec:TTB}
\subsection{{\boldmath$W(W/Z)$} in {\boldmath$\ell\nu$} +HF}
\label{sec:WWZ}
\begin{figure}
\resizebox{0.75\columnwidth}{!}{
  \includegraphics{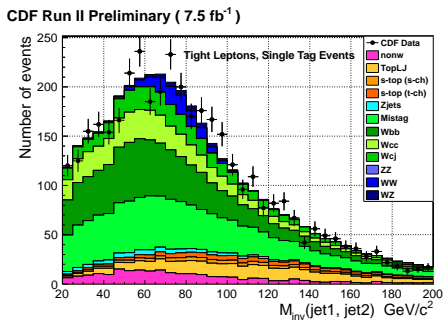} }
\caption{Dijet mass in the 1-tag channel of the CDF search for $W(W/Z)$~\cite{CDFWWZ}.}
\label{CDFWWZ1tag}       
\end{figure}

The CDF collaboration performed a search for $(W\to\ell\nu)$ $(W/Z)$ production, where the 
$(W/Z)$ decays to HF jets~\cite{CDFWWZ}. This analysis uses an integrated luminosity
of 7.5~fb$^{-1}$. The basic selection criteria are: a lepton ($e$ or $\mu$) with 
$p_{\mathrm T} > 20$~GeV, missing $E_{\mathrm T}$ (MET) $> 20$~GeV, and exactly two jets with 
$p_{\mathrm T} > 20$~GeV and $|\eta| < 2$. The bulk of the MJ background is rejected by
an SVM, and the remaining MJ and $W$+jets normalizations are obtained from a template fit
to the MET distribution. Al least one jet is required to be $b$-tagged by a secondary vertex 
algorithm. The final discriminants used are the dijet mass in the 1-tag and 2-tag channels,
of which an example is shown in Fig.~\ref{CDFWWZ1tag}. A signal cross section of $1.1^{+03}_{-04}$ 
times its standard model (SM) expectation is obtained, holding the $WW/WZ$ production ratio to 
its SM value. The significance of the diboson signal is 3.0 standard deviations (s.d.) from the
background-only hypothesis (3.0 expected). In this analysis, however, most of the sensitivity is 
actually coming from $WW$ production in the 1-tag channel, with $W\to cs$.

\begin{figure}
\resizebox{0.75\columnwidth}{!}{
  \includegraphics{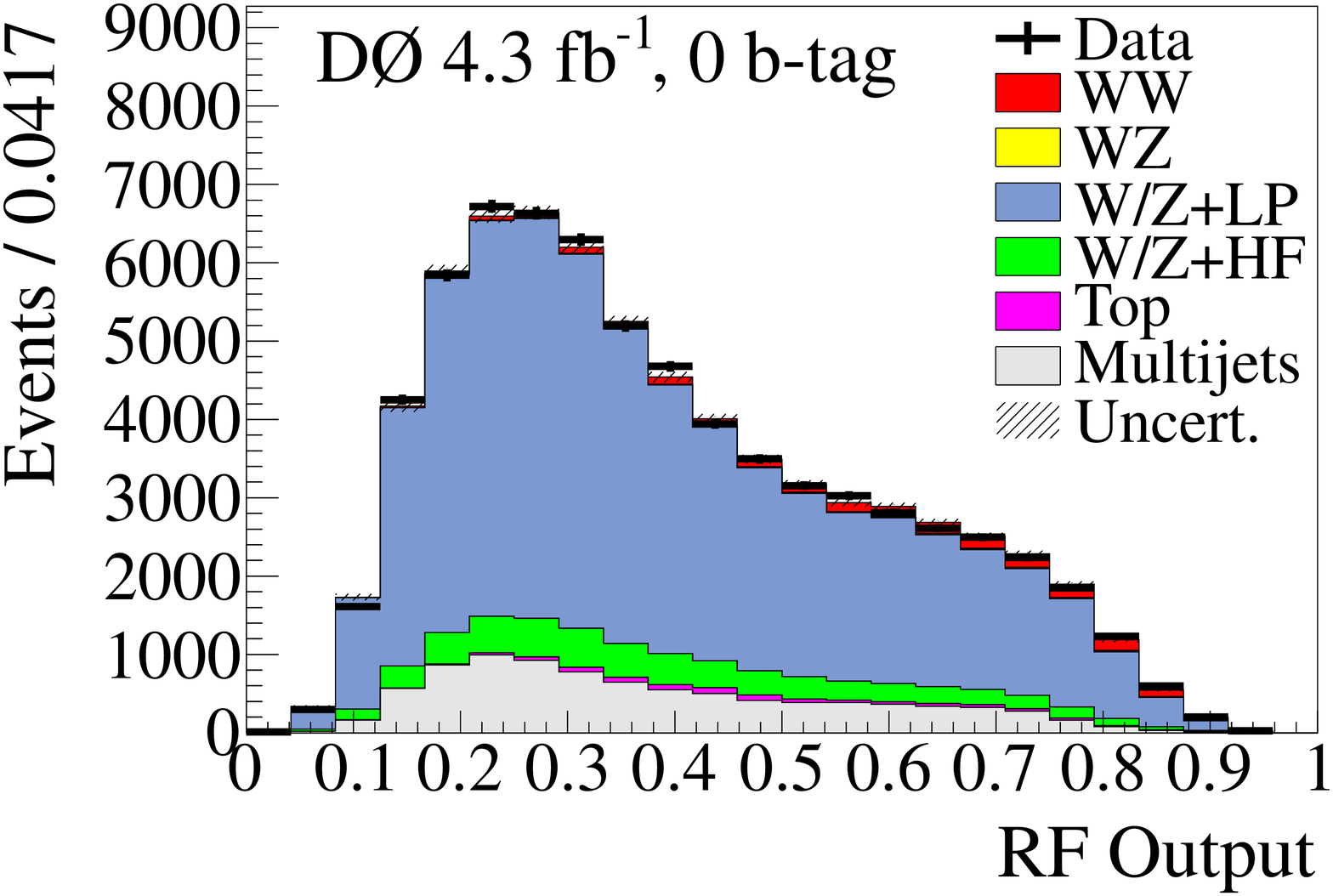} }\\
\resizebox{0.75\columnwidth}{!}{
  \includegraphics{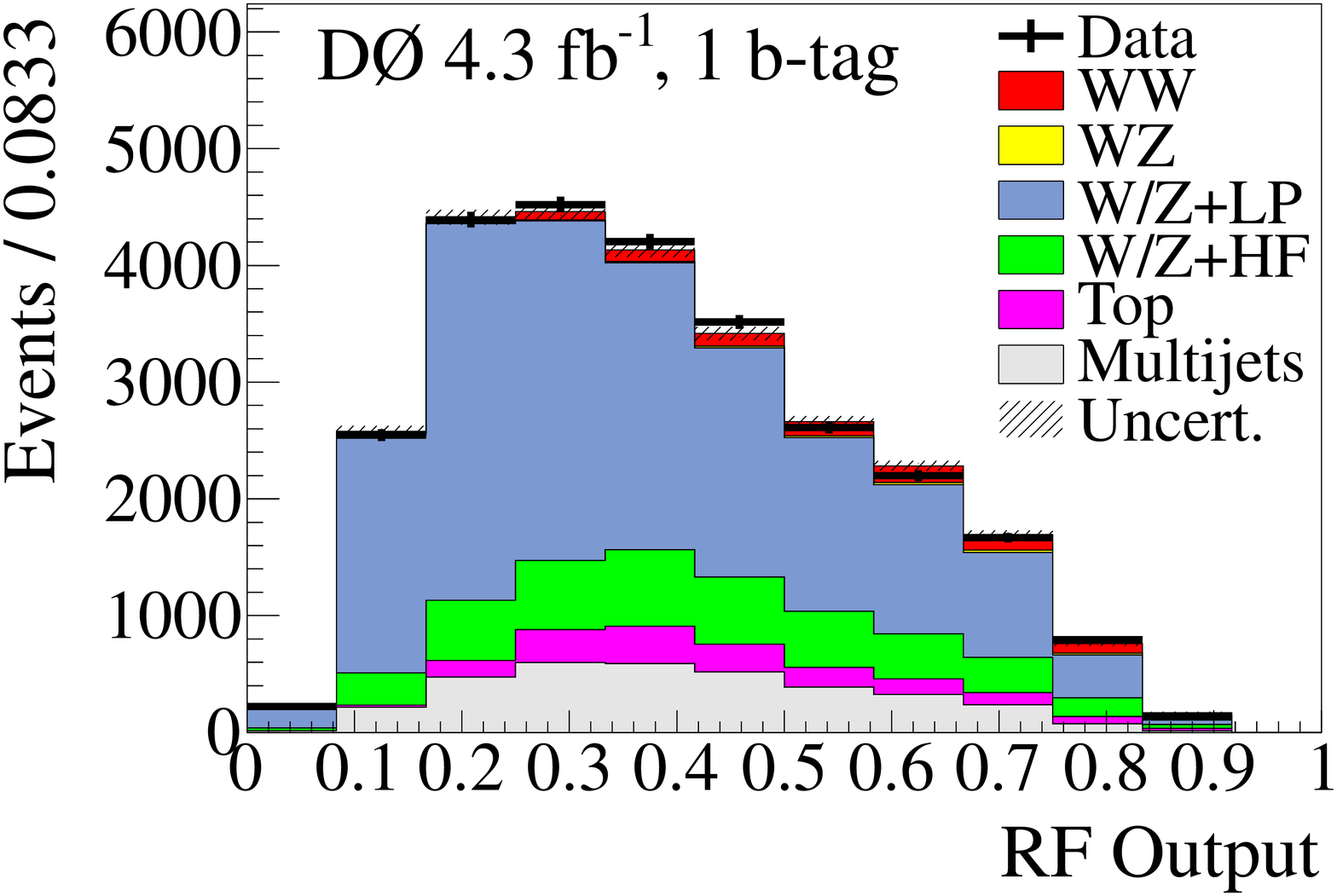} }\\
\resizebox{0.75\columnwidth}{!}{
  \includegraphics{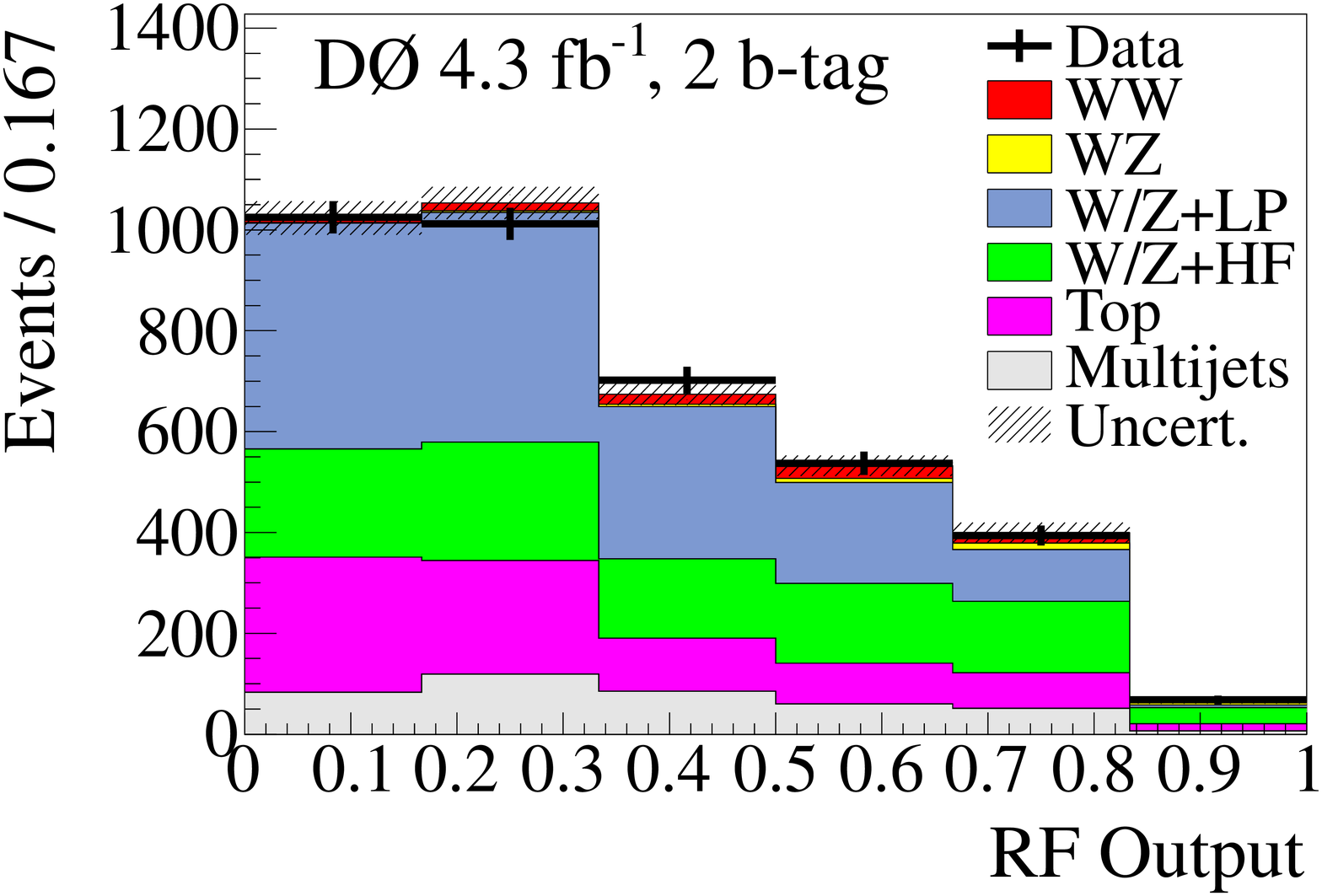} }
\caption{Final discriminant in the 0-, 1-, and 2-tag channels of the D0 search for $W(W/Z)$~\cite{D0WWZ}.}
\label{D0WWZRF}       
\end{figure}

The D0 collaboration performed a search for this same final state~\cite{D0WWZ}, using an integrated
luminosity of 4.3~fb$^{-1}$. The basic selection criteria are: an electron (muon) with 
$p_{\mathrm T} > 20 (15)$~GeV , $\mathrm{MET} > 20$~GeV, and at least two jets with 
$p_{\mathrm T} > 20$~GeV and $|\eta| < 2.5$. To reject most of the MJ background, a so-called 
``triangle cut'' is applied: $m_\mathrm{T} > 40$~GeV~$-0.5$MET, where $m_\mathrm{T}$ is the 
transverse mass of the lepton and the missing $E_\mathrm{T}$. Twelve operating points are defined
for the D0 $b$-tagging algorithm, corresponding to an increasing $b$ purity. The loosest operating
point is chosen to define 0-, 1-, and 2-tag samples, where the two jets with largest $p_\mathrm{T}$ 
(leading jets) are considered for $b$ tagging. But the actual values of the tightest operating 
point passed by each jet are used among the 15 inputs of the final discriminant, a random forest.
The RF outputs in the three $b$-tag channels are shown in Fig.~\ref{D0WWZRF}, from which a $W(W/Z)$
production cross section of $1.2 \pm 0.2$ times its SM expectation is obtained, holding the 
$WW/WZ$ production ratio to its SM value. The signal significance is 8.0 s.d. (6.0 expected). More
relevant for the purpose of this presentation is the result obtained if the $WW$ production cross
section is constrained to its SM value within its uncertainty of 7\%: a cross section of 
$1.3 \pm 0.6$ times its SM expectation for $WZ$ production, and a significance of 2.2 s.d. 
(1.2 expected) for the $WZ$ signal. 

\subsection{{\boldmath$ Z(W/Z)$} in {\boldmath$\ell\ell$} +HF}
\label{sec:ZWZ}
The CDF collaboration performed a search for $(Z\to\ell\ell)$ $(W/Z)$ production, where the 
$(W/Z)$ decays to HF jets~\cite{CDFZWZ}. This analysis uses an integrated luminosity
of 6.6~fb$^{-1}$. The basic selection criteria are: an $ee$ or $\mu\mu$  lepton pair with 
$p_{\mathrm T}(\ell)> 20$~GeV, $76 < m_{\ell\ell} < 106$~GeV, and at least two jets with 
$p_{\mathrm T} > 20$~GeV and $|\eta| < 2$. The originality of this analysis is that it uses, after
$b$-tagging to separate heavy and light flavor jets, a new quark-gluon discriminant to split the 
light-flavor sample into a quark-rich and a gluon-rich sample. The final discriminant is the
dijet mass in each of the three sub-samples so defined. The analysis however does not have (yet)
enough sensitivity for the observation of a diboson signal, and only a 95\% C.L. upper limit of 
1.3 times the SM expectation for $Z(W/Z)$ production has been set (2.3 expected).

\subsection{{\boldmath$(W/Z)Z$} in MET+HF}
\label{sec:WZZ}
\begin{figure}
\begin{center}
\resizebox{0.75\columnwidth}{!}{
  \includegraphics{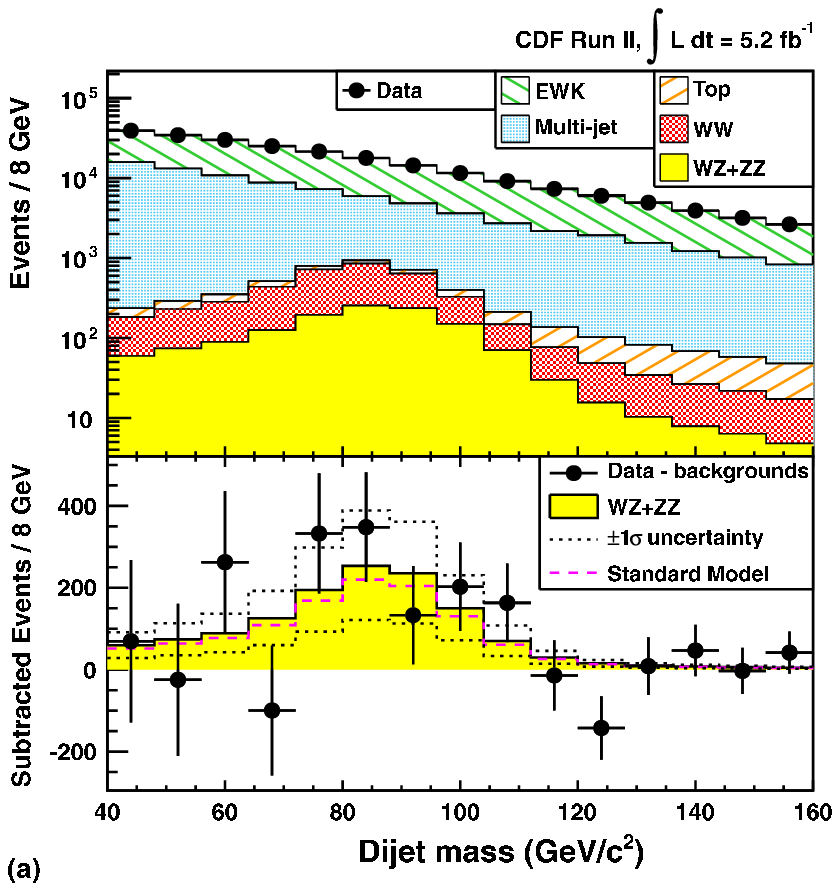}} \\
\resizebox{0.75\columnwidth}{!}{
  \includegraphics{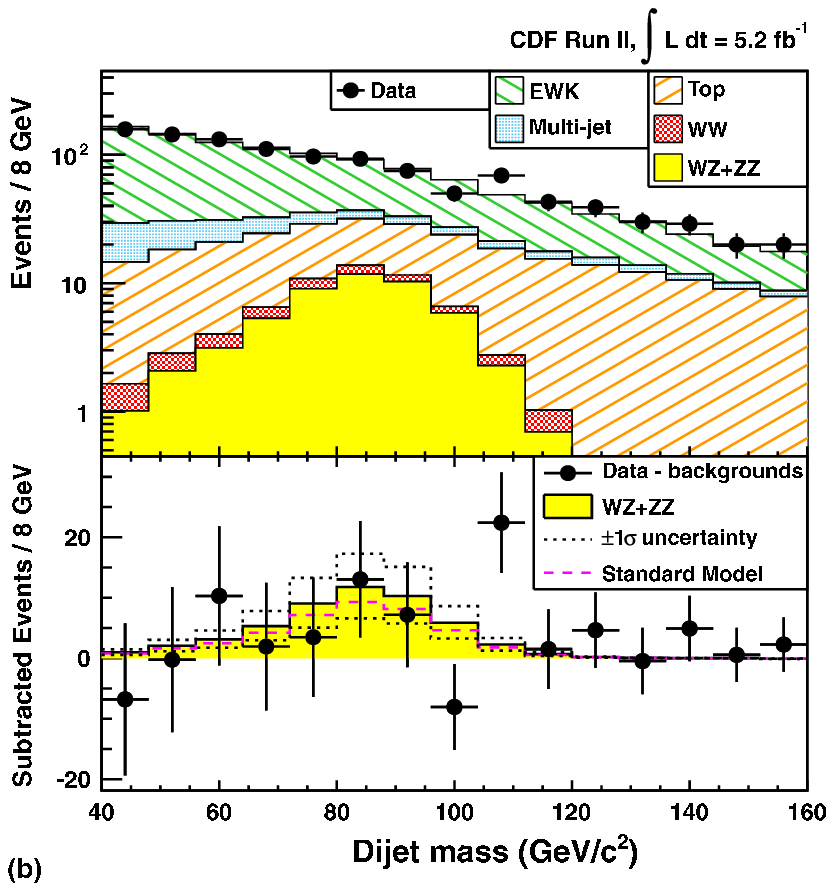}} \\
\end{center}
\caption{Dijet mass in the (0+1)- and 2-tag channels
of the CDF search for $(W/Z)Z$~\cite{CDFWZZ}.}
\label{CDFWZZ12tag}       
\end{figure}

The CDF collaboration performed a search for $(W\to\ell\nu$/ $Z\to\nu\nu)Z$ production, where the 
$Z$ decays to HF jets~\cite{CDFWZZ}. This analysis uses an integrated luminosity
of 5.2~fb$^{-1}$. The basic selection criteria are: $\mathrm{MET} > 50$~GeV, supplemented with a
requirement on the missing $E_{\mathrm T}$ significance, at most one lepton ($e$ or $\mu$), and 
at least two jets with $p_{\mathrm T} > 20$~GeV and $|\eta| < 2$. The bulk of the MJ background is 
rejected by the requirement that the azimuthal angle between the missing $E_{\mathrm T}$ and any jet
be larger than 0.4 radians. 
The shape of the remaining MJ background is taken from events in which the azimuthal angle between
the missing $E_{\mathrm T}$ and the missing $p_{\mathrm T}$, the former calculated from calorimeter 
information and the latter from charged particle tracks, is larger than one radian.
The remaining sample is divided into two sub-samples, one with zero 
or one $b$-tagged jet, and the other with two $b$-tagged jets. 
The final discriminant is the dijet mass in each of those two sub-samples, 
as shown in Fig.~\ref{CDFWZZ12tag}. The results are 
obtained with the $WW$ production cross section fixed to its SM value, within its uncertainty,
and with the $WZ/ZZ$ ratio fixed to its SM value.
The normalization of the $(W/Z)$+jets background is allowed to float independently in the
two sub-samples. A signal cross section of $1.1^{+07}_{-06}$ 
times its standard model expectation is obtained. 
The significance of the diboson signal is 1.9 s.d. (1.7 expected). 
Because this analysis accepts events with zero or one
lepton, both $WZ$ and $ZZ$ production contribute to the sensitivity.

\section{The ultimate benchmark}
\label{sec:TB}
The following mutually exclusive analyses by the D0 collaboration
are exact copies of the corresponding searches for a
low-mass Higgs boson. The only changes are the MVA trainings, in which the signal is now $WZ/ZZ$
instead of $WH/ZH$, while $WW$ remains a background. 

In each of the analyses, fits of the nuisance parameters were
performed to the final discriminant outputs in all sub-channels, with systematic uncertainties
correlated across signal and backgrounds as appropriate. The main sources
of uncertainty are: the ratio of heavy to light flavor production in $(W/Z)$+jets; the various
object reconstruction and identification efficiencies; the jet energy calibration and resolution;
the $b$-tagging efficiency and the rate of wrongly tagged light-flavor jets. Three kinds of fits
to the data 
were performed: in one of them, the signal rate was also fitted, and the diboson production cross 
section was therefore measured; in the other ones, the signal rate was set either to zero
(background-only hypothesis) or to its SM value (signal+background hypothesis), from which the
observed LLR was deduced. These fits were repeated on pseudo-experiments, and the fraction of 
background-only pseudo-experiments yielding a cross section value at least as large as 
the one observed 
(p-value) was used to assess the signal significance, next translated into Gaussian standard 
deviations. The consistency of the result with the standard model was determined
in a similar way, using signal+background pseudo-experiments. Unless
otherwise specified, the ratio of the $WZ$ and $ZZ$ production cross sections was held fixed to
its SM value. 

\subsection{{\boldmath$WZ$} in {\boldmath$\ell\nu$}+HF}
\label{sec:WZ}
\begin{figure}
\resizebox{0.75\columnwidth}{!}{
  \includegraphics{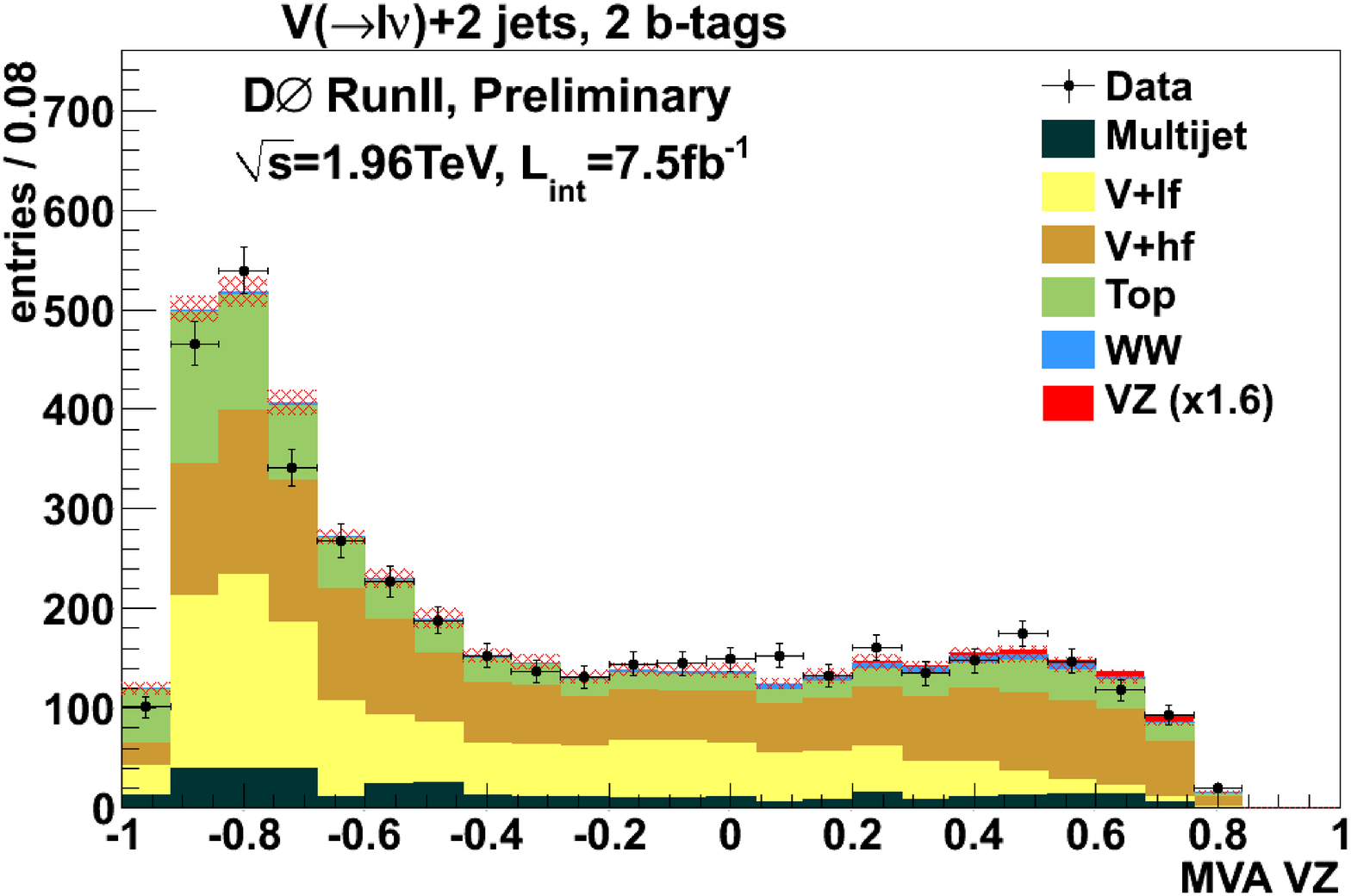} } \\
\resizebox{0.75\columnwidth}{!}{
  \includegraphics{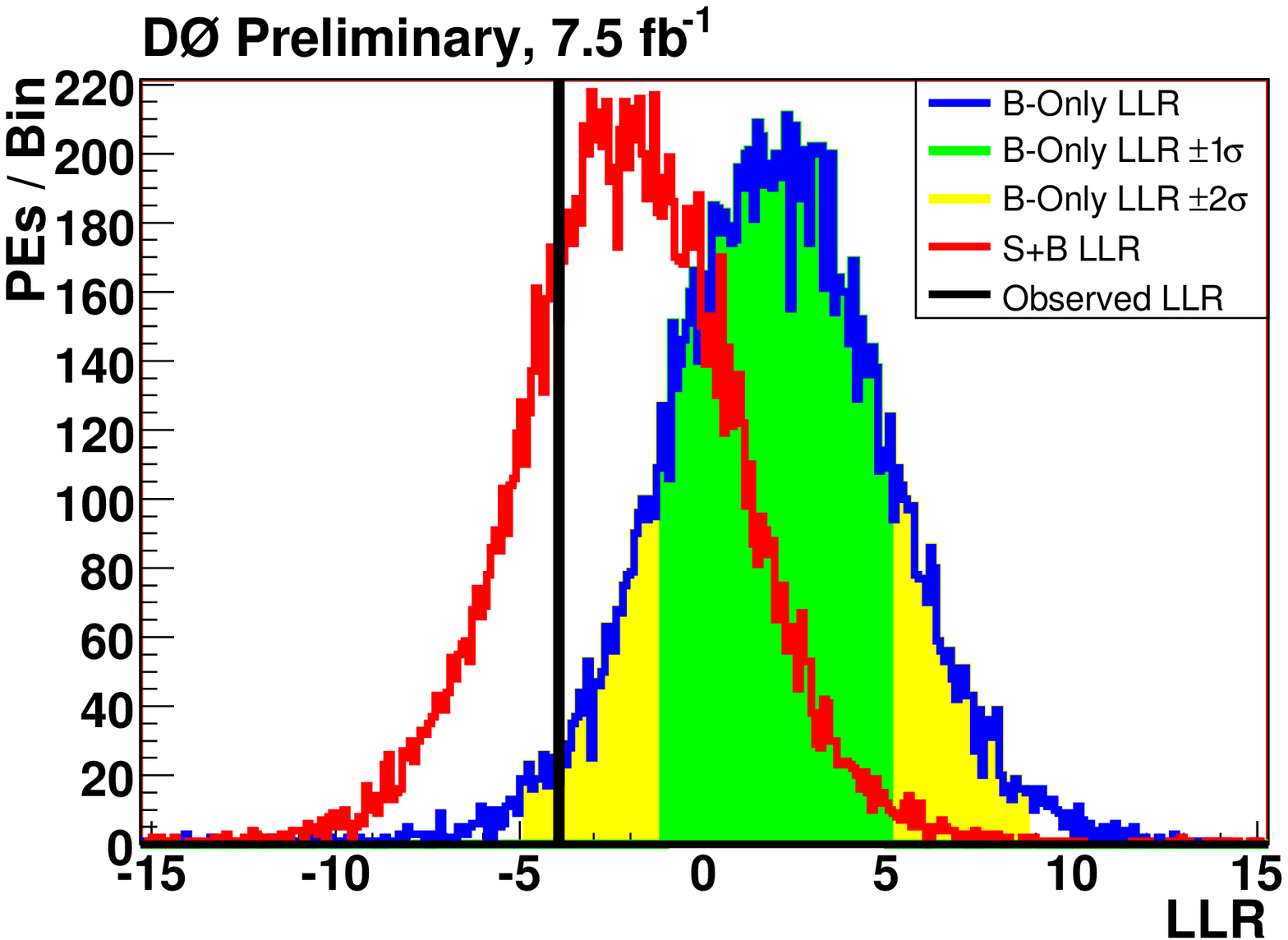} }
\caption{In the D0 search for $WZ$~\cite{D0lvbb}: final discriminant in the 2-tag channel (top); observed LLR 
value and expected LLR distributions in the background-only and signal+background hypotheses (bottom).} 
\label{D0lvbbBDT}       
\end{figure}

The D0 collaboration performed this search in a data sample corresponding to an integrated 
luminosity of 7.5~fb$^{-1}$\cite{D0lvbb}. An isolated lepton ($e$ or $\mu$), missing $E_{\mathrm T}$,
and two or three jets are required. The MJ background is rejected by a ``triangle cut'' in the
muon channel and using a BDT in the electron channel. The sample is split into a 1-tag and a 2-tag
sub-sample, based on the loosest $b$-tagging operating point as in the D0 analysis reported in
Sec.~\ref{sec:WWZ}. As in that same analysis, the remaining $b$-tagging information is used in the
final discriminant, a BDT with 14 inputs. The BDT output in the 2-tag channel is shown in 
Fig.~\ref{D0lvbbBDT} (top). A signal cross section of $1.6 \pm 0.8$ times its standard model 
expectation is obtained, with a significance of 2.2 s.d. (1.4 expected). 
The observed LLR is compared to expected distributions in the background-only and signal+background
hypotheses in Fig.~\ref{D0lvbbBDT} (bottom). 
The sensitivity of this search is dominated by $WZ$ production.

\subsection{{\boldmath$ZZ$} in {\boldmath$\ell\ell$}+HF}
\label{sec:ZZ}
\begin{figure}
\resizebox{0.75\columnwidth}{!}{
  \includegraphics{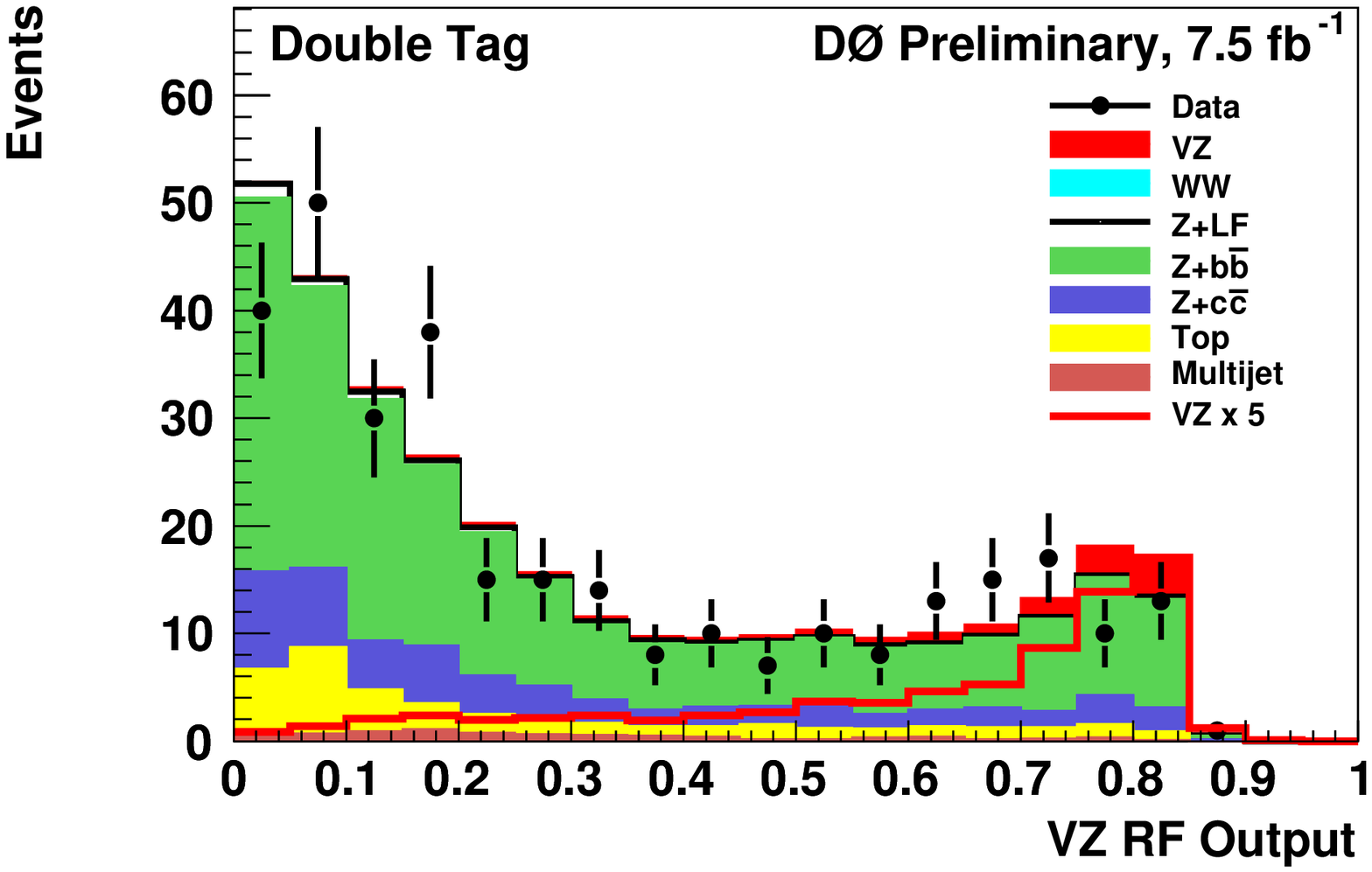} } \\
\resizebox{0.75\columnwidth}{!}{
  \includegraphics{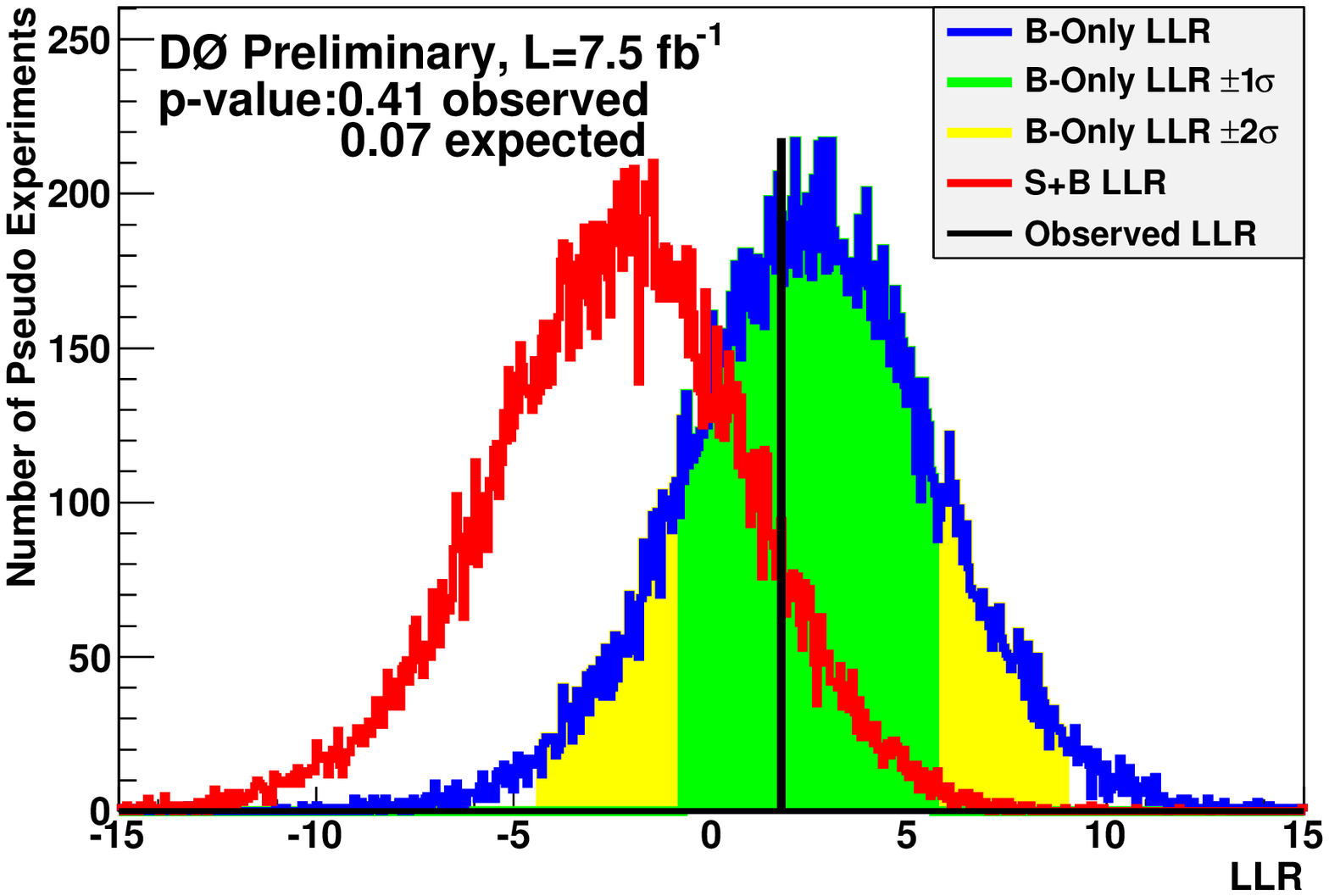} }
\caption{In the D0 search for $ZZ$~\cite{D0llbb}: final discriminant in the 2-tag channel (top); observed LLR 
value and expected LLR distributions in the background-only and signal+background hypotheses (bottom).} 
\label{D0llbbBDT}       
\end{figure}

The D0 collaboration performed this search in a data sample corresponding to an integrated 
luminosity of 7.5~fb$^{-1}$\cite{D0llbb}. An electron or muon pair in a $Z$-mass window ,
and two or three jets are required. The sample is split into a 1-tag and a 2-tag
sub-sample: in the 2-tag sample, two jets are $b$-tagged, one tightly and one loosely; 
in the 1-tag sample, one of the jets is tightly $b$-tagged, with no other jet passing the loose 
$b$-tag requirement. To take advantage of the absence of missing $E_{\mathrm T}$ in the signal, 
a kinematic fit is performed, which improves significantly the dijet mass resolution. 
The final discriminant is an RF with 19
inputs, the output of which is  shown for the 2-tag channel in 
Fig.~\ref{D0llbbBDT} (top). A signal cross section of $0.1 \pm 0.6$ times its standard model 
expectation is obtained. The significance is only 0.1 s.d. (1.5 expected), not inconsistent
with the signal+back\-ground hypothesis.
The observed LLR is compared to expected distributions in the background-only and signal+
background hypotheses in 
Fig.~\ref{D0llbbBDT} (bottom). The sensitivity of this search is dominated by $ZZ$ production.

\begin{figure}
\resizebox{0.75\columnwidth}{!}{
  \includegraphics{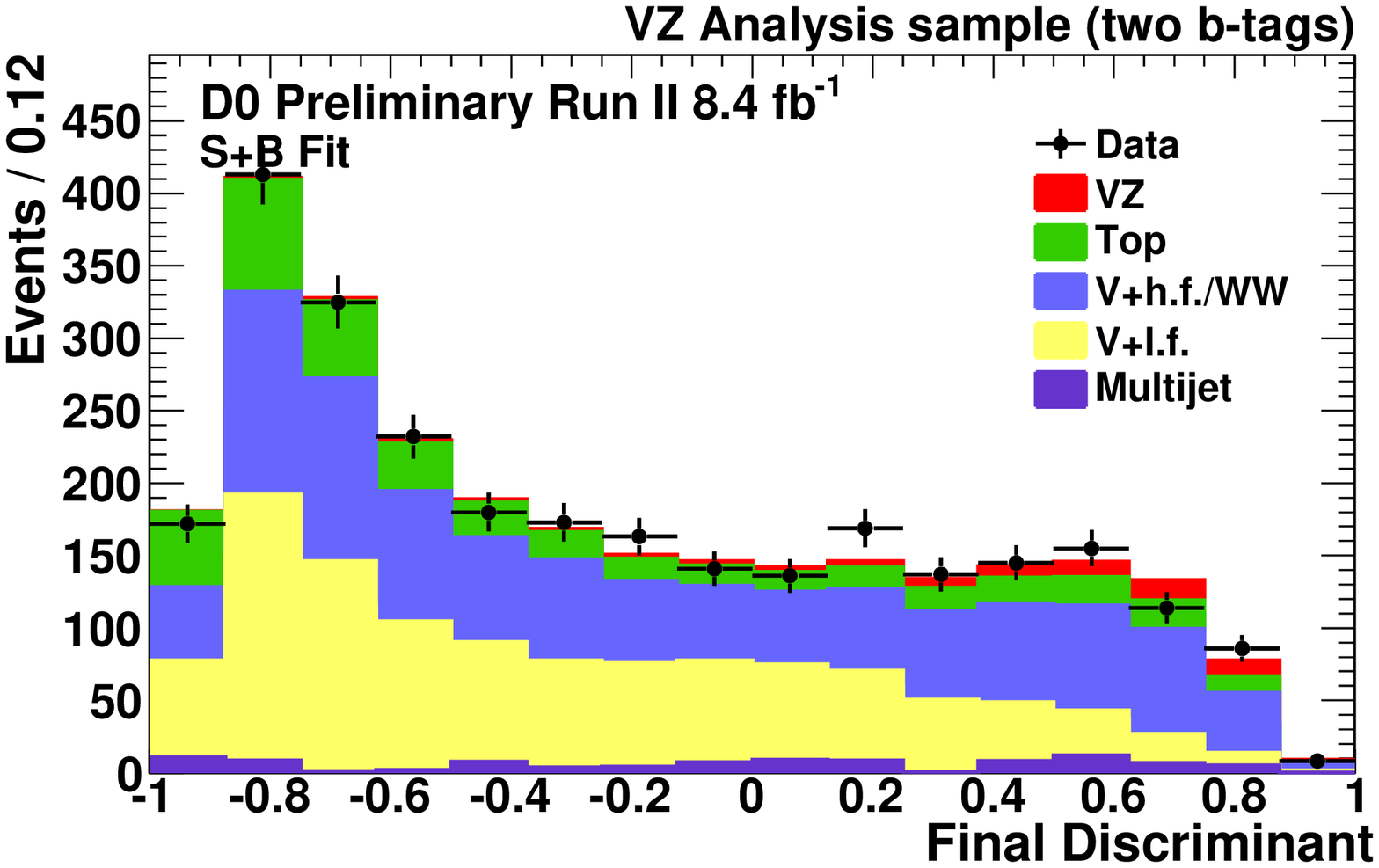} } \\
\resizebox{0.75\columnwidth}{!}{
  \includegraphics{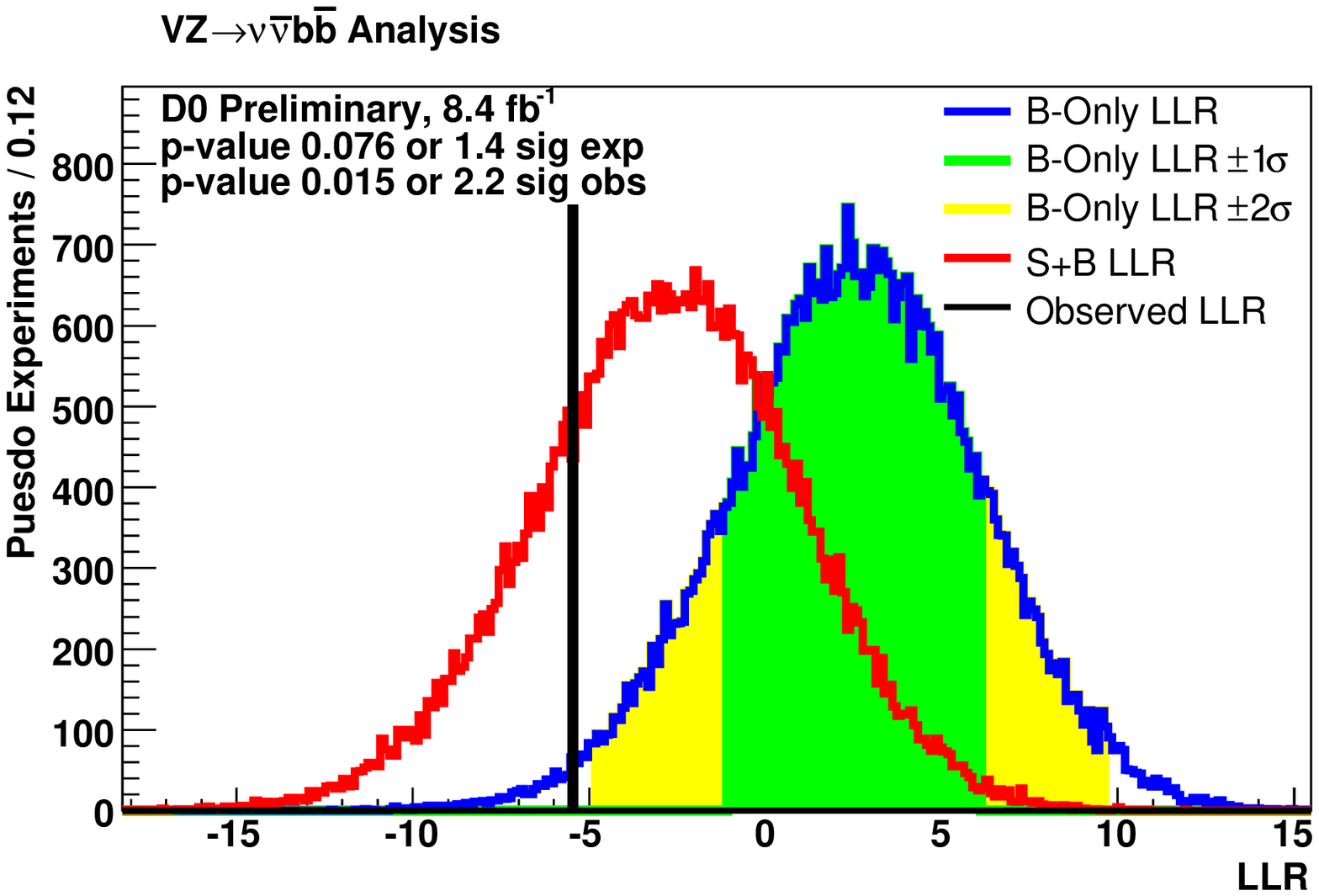} }
\caption{In the D0 search for $(Z/W)Z$~\cite{D0vvbb}: final discriminant in the 2-tag channel (top); 
observed LLR value and expected LLR distributions in the background-only and signal+background 
hypotheses (bottom).} 
\label{D0vvbbBDT}       
\end{figure}

\subsection{{\boldmath$(Z/W)Z$} in MET+HF}
\label{sec:VZ}
The D0 collaboration performed this search in a data sample corresponding to an integrated 
luminosity of 8.4~fb$^{-1}$\cite{D0vvbb}. The selection requires  a large MET with a large
significance, and two jets not back-to-back in the plane transverse to the beam direction; it 
rejects events with an electron or a muon satisfying the criteria of the WZ search reported
in Sec.~\ref{sec:WZ}. The bulk of the MJ background is next rejected using a BDT. The sample is 
split into a 1-tag and a 2-tag sub-sample, based on the loosest $b$-tagging operating point as in 
the D0 analysis reported in Sec.~\ref{sec:WWZ}. As in that same analysis, the remaining $b$-tagging
information is used in the final discriminant, a BDT with 32 (!) inputs. The BDT output in the 
2-tag channel is shown in Fig.~\ref{D0vvbbBDT} (top). A signal cross section of $1.5 \pm 0.5$ times
its standard model expectation is obtained, with a significance of 2.8 s.d. (1.9 expected).
The observed LLR is compared to expected distributions in the background-only and signal+background
hypotheses in Fig.~\ref{D0vvbbBDT} (bottom). The sensitivity of this search is shared by 
$(Z\to\nu\nu)Z$ and by $(W\to\ell\nu)Z$ production, where the lepton from the $W$ decay falls 
outside of the acceptance or fails the identification criteria. 

\subsection{Combination}
\label{sec:Combi}
\begin{figure}
\resizebox{0.75\columnwidth}{!}{
  \includegraphics{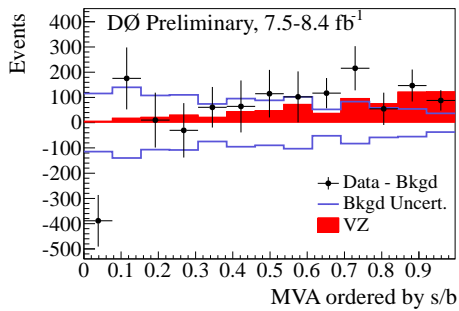} }
\caption{In the combination of D0 searches for $WZ/ZZ$~\cite{D0comb}, final discriminant re-cast 
into common bins of
signal-to-background ratio. The fitted background has been subtracted, with the blue lines
indicating the fitted background uncertainty. The fitted signal is shown as a red histogram.} 
\label{D0MVA}       
\end{figure}

\begin{figure}
\resizebox{0.75\columnwidth}{!}{
  \includegraphics{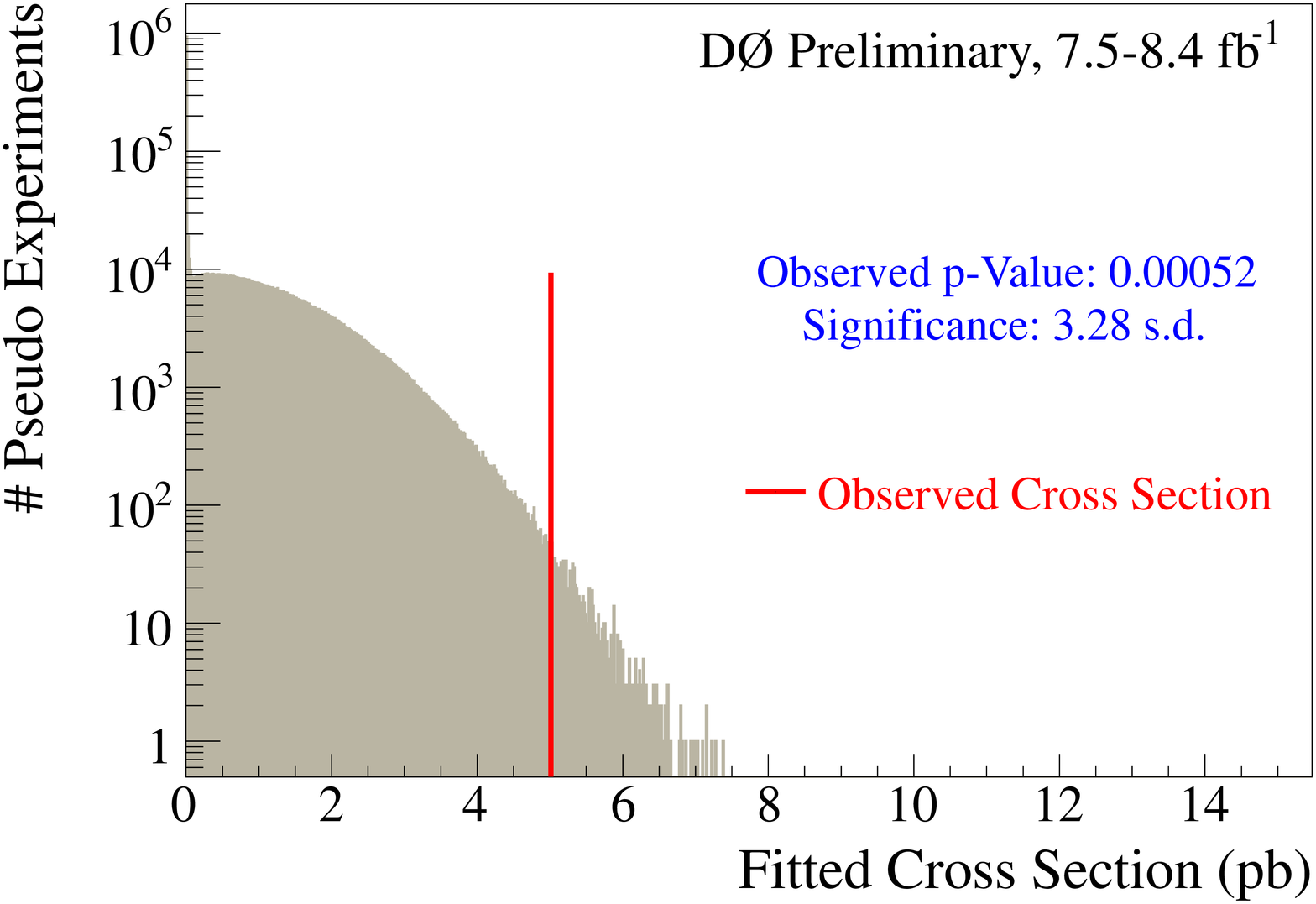} } \\
\resizebox{0.75\columnwidth}{!}{
  \includegraphics{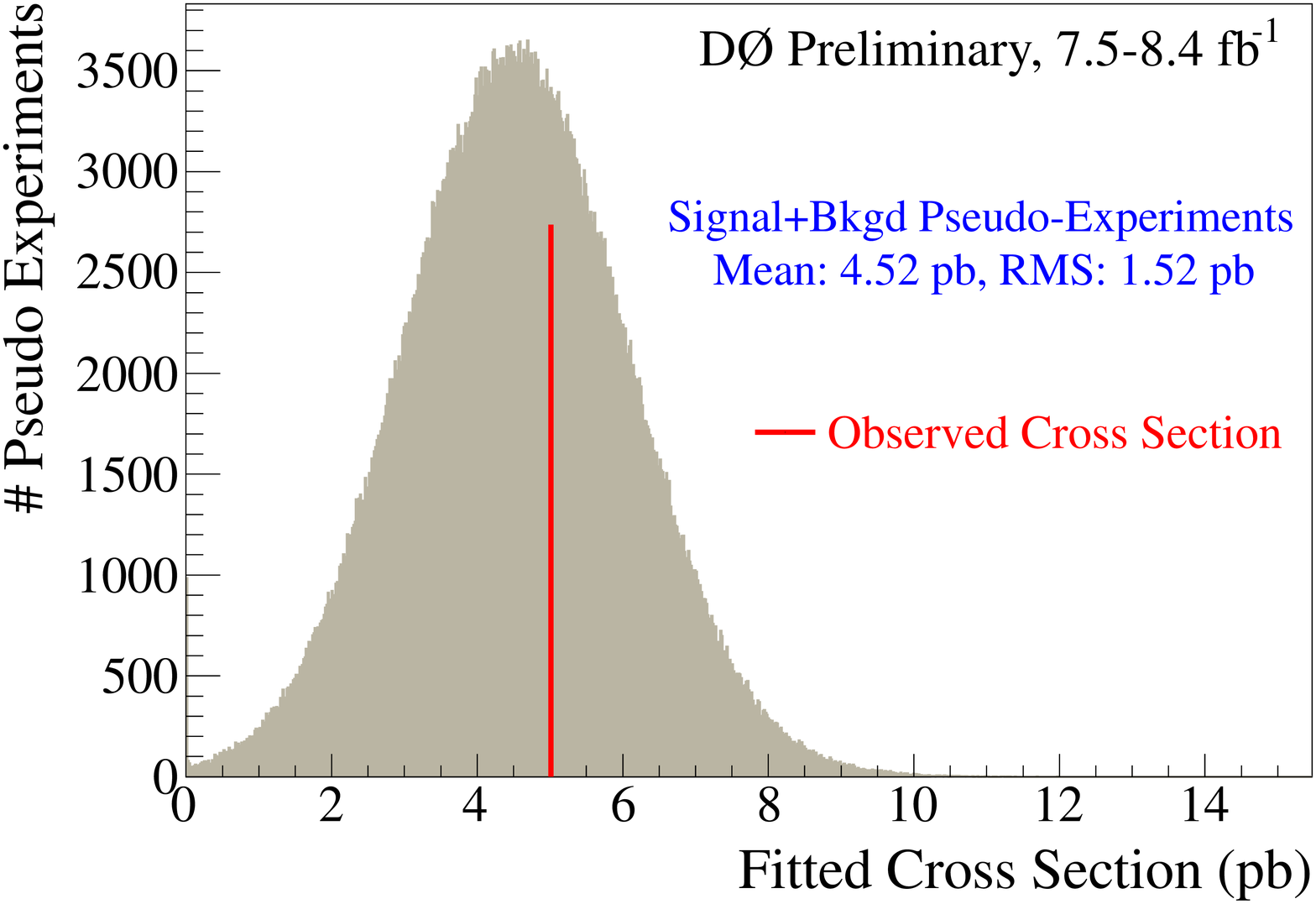} }
\caption{In the combination of D0 searches for $WZ/ZZ$~\cite{D0comb}, comparison of the measured signal cross
section with expectations from pseudo-experiments in the background-only (top) and 
signal+background (bottom) hypotheses.} 
\label{D0pseudo}       
\end{figure}

\begin{figure}
\resizebox{0.75\columnwidth}{!}{
  \includegraphics{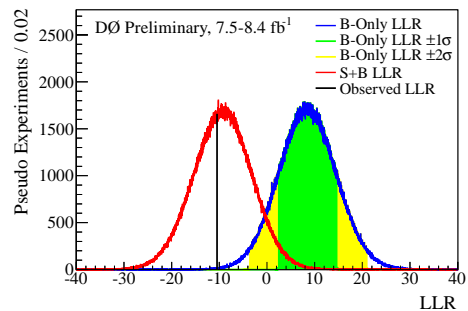} }
\caption{In the combination of D0 searches for $WZ/ZZ$~\cite{D0comb}, observed LLR and expected LLR 
distributions in the signal+background and background-only hypotheses.} 
\label{D0LLR}       
\end{figure}

\begin{figure}
\resizebox{0.75\columnwidth}{!}{
  \includegraphics{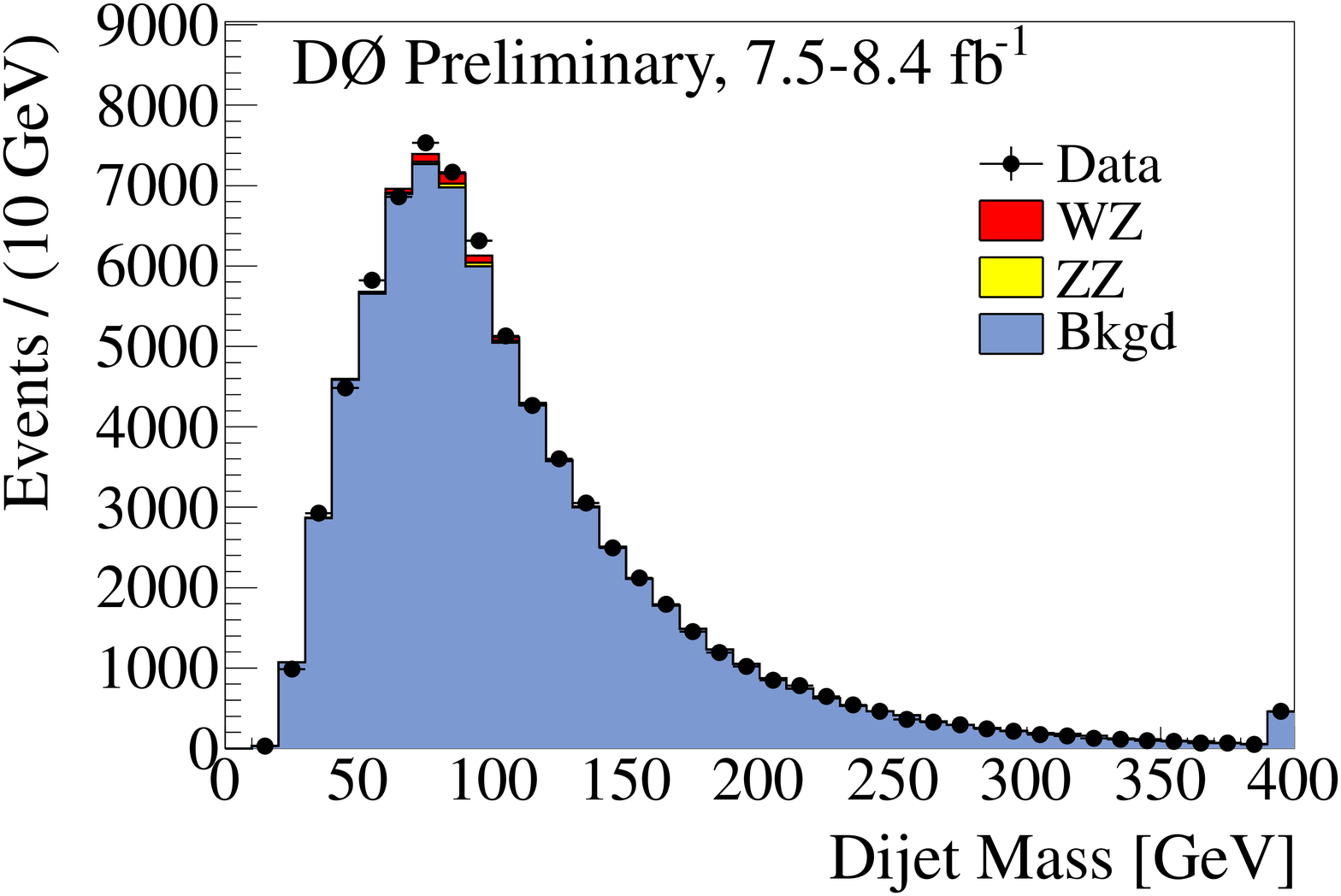} } \\
\resizebox{0.75\columnwidth}{!}{
  \includegraphics{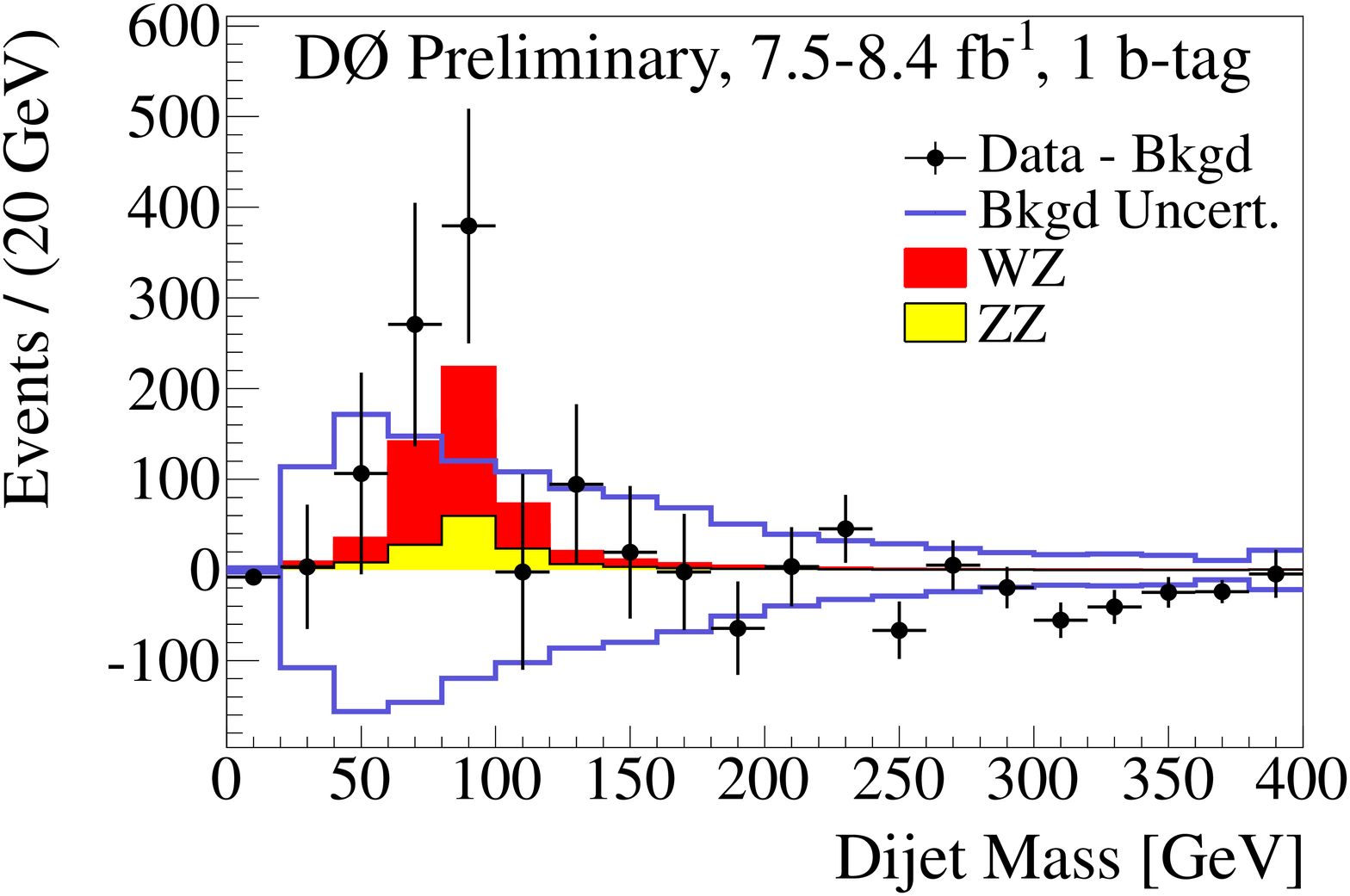} } \\
\resizebox{0.75\columnwidth}{!}{
  \includegraphics{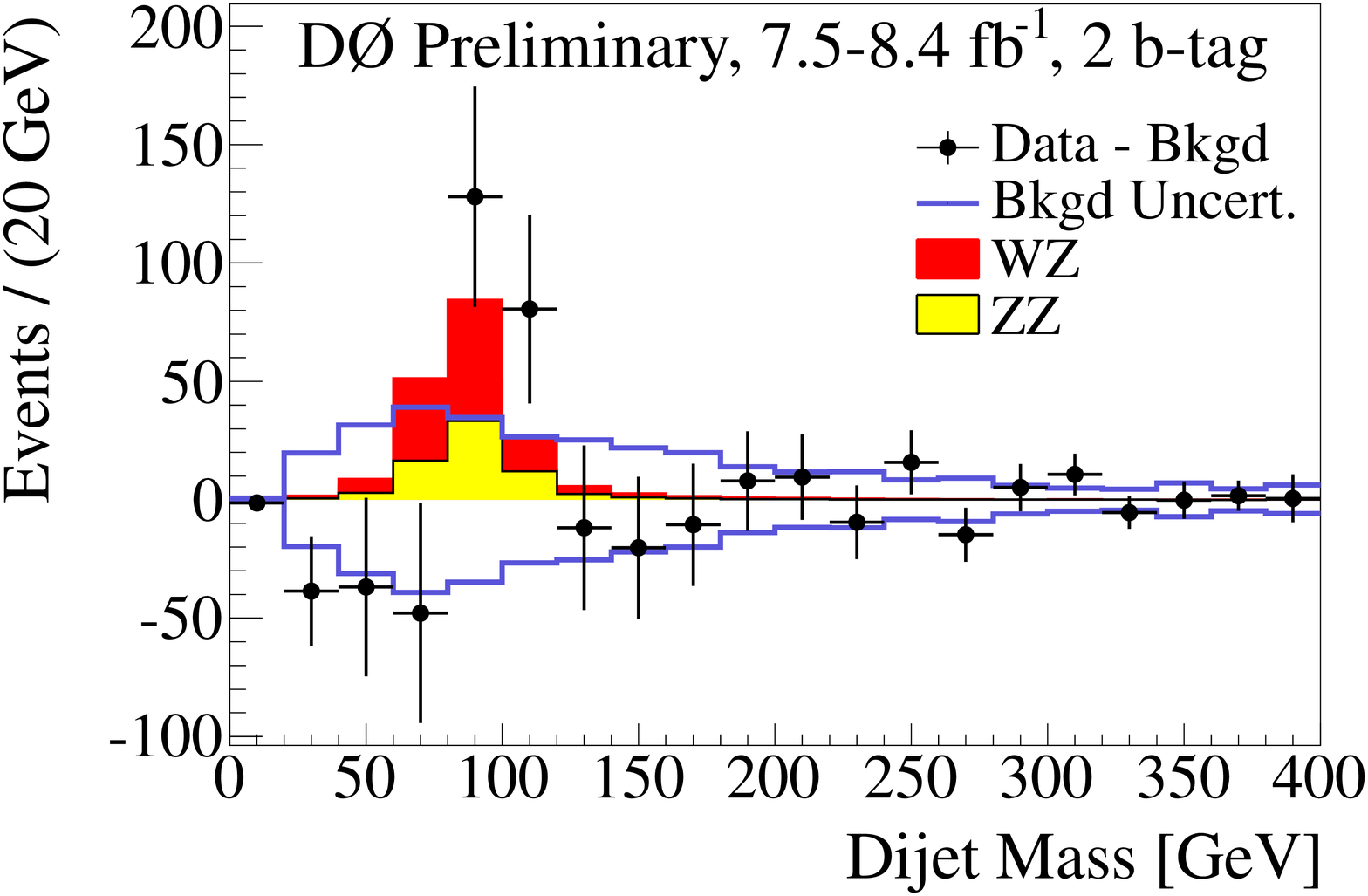} } \\
\resizebox{0.75\columnwidth}{!}{
  \includegraphics{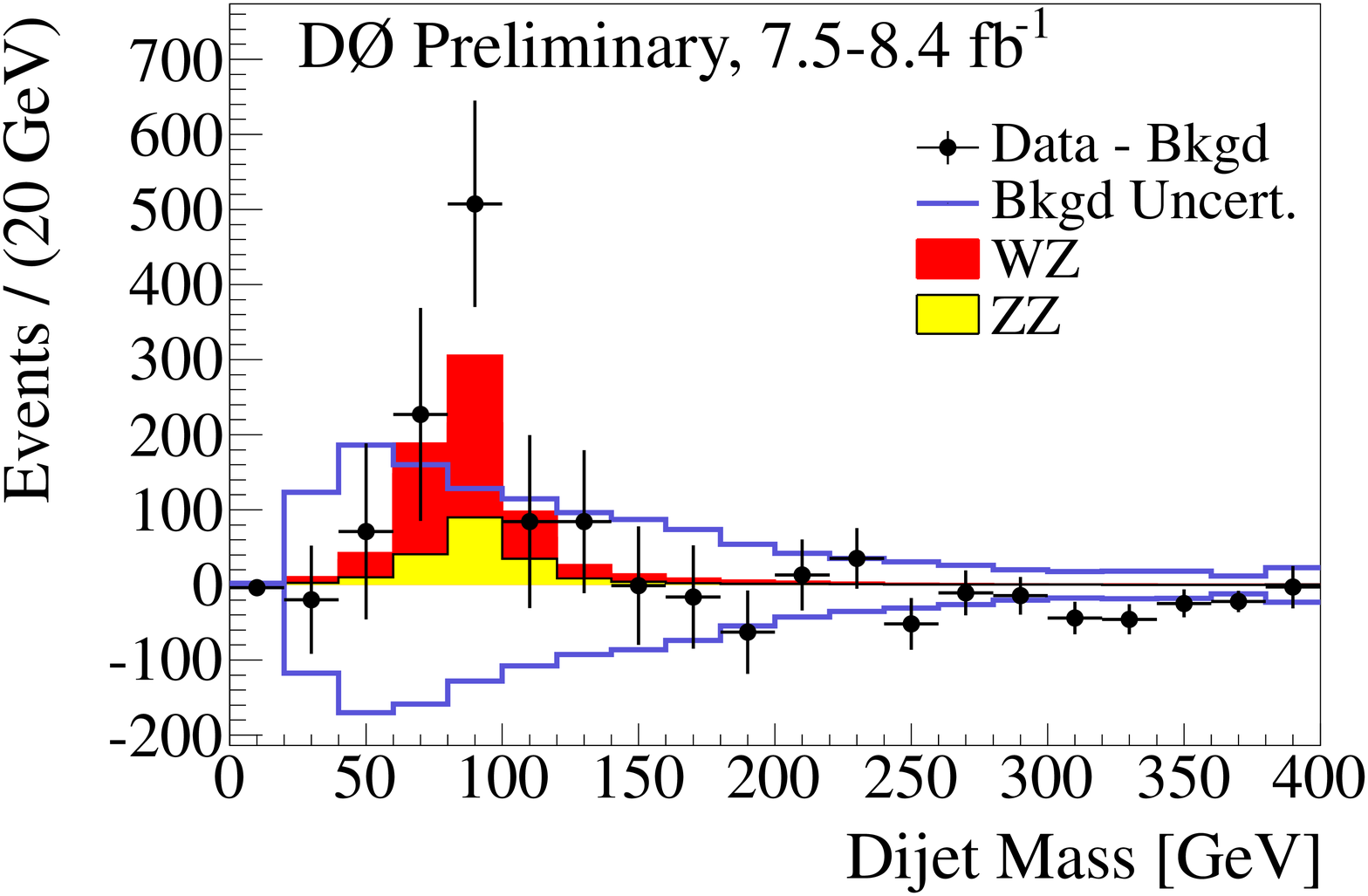} }
\caption{In the combination of D0 searches for $WZ/ZZ$~\cite{D0comb}, post-fit dijet mass distribution in the
1- and 2-tag channels combined (top), and, after fitted background subtraction, in the 1-tag 
channel, 2-tag channel,
and 1- and 2-tag channels combined, from row 2 to 4. The blue lines indicate the fitted background
uncertainties.}
\label{D0mass}       
\end{figure}

The D0 searches for $WZ/ZZ$ reported in the previous subsections were combined using the exact
same techniques as for the combination of the searches for the Higgs boson. 
Since the binning of the final discriminant outputs are not identical in the various sub-channels,
these outputs were re-cast into common bins of signal-to-background ratio. 
The result of the fit from 
which the signal cross section is measured is shown in Fig.~\ref{D0MVA}. The result is 
$1.13 \pm 0.36$ times the SM cross section. This result can be compared with the expectation
from pseudo-experiments drawn in the background-only and signal+background hypotheses in 
Fig.~\ref{D0pseudo}. From the former, a signal significance of 3.3 s.d. is deduced (2.9 expected),
while the latter shows consistency with the signal+background hypothesis within 0.3 s.d. The 
observed LLR is compared to expected distributions in the two hypotheses in Fig.~\ref{D0LLR}.
The results of the fit of the diboson production cross section to the combined final discriminant 
can also be used to plot other quantities, such as the dijet mass, as shown in Fig.~\ref{D0mass}.

\begin{figure}
\resizebox{0.75\columnwidth}{!}{
  \includegraphics{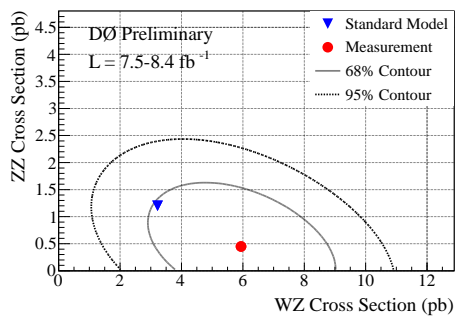} }
\caption{In the combination of D0 searches for $WZ/ZZ$~\cite{D0comb}, fitted $WZ$ and $ZZ$ production cross
sections compared to the SM expectation.} 
\label{D02D}       
\end{figure}

A fit was also performed in which the $WZ$ and $ZZ$ cross sections were left uncorrelated. The
results are, relative to the SM cross sections, $1.8 \pm 0.5$ for $WZ$ and $0.4 \pm 1.1$ for $ZZ$. 
These results are correlated as shown in Fig.~\ref{D02D}, where it can be seen that 
the SM expectation lies within the 68\% C.L. contour of the data.
The deviation from the SM is as expected, given the
results of the individual channels reported in the previous subsections. 

\subsection{A frequently asked question}
\label{sec:FAQ}
The question was raised at this conference of the flavor composition of the final diboson samples.
The answer is given here for the D0 search in the MET+HF final state~\cite{D0vvbb}.

In the 1-tag channel:
\begin{itemize}
\item $bb$: 16\%
\item $cc$: 19\%
\item $cs$: 23\%
\item other: 41\%
\end{itemize}

In the 2-tag channel:
\begin{itemize}
\item $bb$: 60\%
\item $cc$: 22\%
\item $cs$: 8\%
\item other: 9\%
\end{itemize}

It should however be kept in mind that these values correspond to a definition of the 1-tag and 
2-tag channels according to the loosest of the twelve $b$-tagging operating points used at D0, 
and that the remaining $b$-tagging information is used as input to the final discriminant. 
The weight of the $bb$ component is therefore substantially larger
than the above numbers would suggest.

\section{Summary}
\label{sec:Final}
A number of analyses have recently addressed the search for diboson production at the Tevatron
with heavy-flavor (HF) jets in the final state. The main results are the following.

\begin{itemize}
\item The CDF collaboration observed $WW/WZ$ production in the $\ell\nu$+HF channel 
with a 3.0 standard deviation (s.d.) significance~\cite{CDFWWZ}. 
\item In that same channel, the D0 collaboration reached a significance of 2.2 s.d. for
$WZ$ production alone~\cite{D0WWZ}.
\item The CDF collaboration obtained a significance of 1.9 s.d. for $WZ/ZZ$ production 
in the missing $E_{\mathrm T}$(MET)+HF channel, with zero or one lepton accepted~\cite{CDFWZZ}.
\end{itemize}
The D0 collaboration recycled their searches for a low-mass Higgs boson in data samples
corresponding to integrated luminosities of 7.5 to 8.4~fb$^{-1}$, using $WZ$ and $ZZ$ as a 
signal instead of $WH$ and $ZH$. 
\begin{itemize}
\item For the $WZ$ signal in the $\ell\nu$+HF final state, a significance of 2.2 s.d. 
was obtained~\cite{D0lvbb}.
\item For the $ZZ$ signal in the $\ell\ell$+HF final state, the significance is only 
0.1 s.d.~\cite{D0llbb}.
\item For the $WZ/ZZ$ signal in the MET+HF final state with no identified leptons, a 
significance of 2.8 s.d. was obtained~\cite{D0vvbb}.
\end{itemize}
Those three searches were combined to reach a significance of 3.3 s.d. (2.9 expected), 
thereby establishing evidence for diboson production in final states containing heavy-flavor 
jets~\cite{D0comb}. 
The production cross section was measured to be $1.13 \pm 0.36$ times its standard model 
expectation.

These analyses have provided a direct validation of the procedures and techniques used in the 
searches for a low-mass Higgs boson at the Teavtron.


\begin{thebibliography}{}
\bibitem{Robson}
A. Robson, these proceedings.
\bibitem{Sforza}
F. Sforza, these proceedings.
\bibitem{MCFM}
J. M. Campbell and R. K. Ellis, Phys. Rev. D \textbf{60}, (1999) 113006.
\bibitem{CDFWWZ}
The CDF Collaboration, CDF Note 10598.
\bibitem{D0WWZ}
The D0 Collaboration, submitted to Phys. Rev. Lett., arXiv:1112.0536 [hep-ex].
\bibitem{CDFZWZ}
The CDF Collaboration, CDF Note 10601.
\bibitem{CDFWZZ}
The CDF Collaboration, to appear in Phys. Rev. D, arXiv:1108.2060 [hep-ex].
\bibitem{D0lvbb}
The D0 Collaboration, D0 Note 6220-CONF.
\bibitem{D0llbb}
The D0 Collaboration, D0 Note 6256-CONF.
\bibitem{D0vvbb}
The D0 Collaboration, D0 Note 6223-CONF.
\bibitem{D0comb}
The D0 Collaboration, D0 Note 6260-CONF.
\end{thebibliography}
\end{document}